\documentclass[12pt]{article}
\usepackage{amssymb}
\usepackage{epsfig}

\catcode`\@=11
\def\numberbysection{\@addtoreset{equation}{section}
        \def\theequation{\thesection.\arabic{equation}}}

\def\beq{\begin{equation}}
\def\eeq{\end{equation}}
\numberbysection
\begin{document}
\begin{titlepage}
\begin{center}
\hfill DFF  1/3/08 \\
\vskip 1.in {\Large \bf Extended BRS symmetry in topological field
theories } \vskip 0.5in P. Valtancoli
\\[.2in]
{\em Dipartimento di Fisica, Polo Scientifico Universit\'a di Firenze \\
and INFN, Sezione di Firenze (Italy)\\
Via G. Sansone 1, 50019 Sesto Fiorentino, Italy}
\end{center}
\vskip .5in
\begin{abstract}
A class of topological field theories like the $BF$ model and
Chern-Simons theory, when quantized in the Landau gauge, enjoys
the property of invariance under a vector supersymmetry, which is
responsible for their finiteness. We introduce a new type of gauge
fixing which makes these theories invariant under an extended
$BRS$ symmetry, containing a new type of field, the ghost of
diffeomorphisms. The presence of such an extension is naturally
related to the vector supersymmetry discussed before.
\end{abstract}
\medskip
\end{titlepage}
\pagenumbering{arabic}
\section{Introduction}

The $BF$ model like the Chern-Simons theory are classically a
class of topological models of Schwarz type \cite{1}. Their
topological nature  \cite{2}-\cite{3} is ensured by the property
of invariance not only under the $BRS$ symmetry but also under a
vector supersymmetry \cite{4} making these models finite
\cite{5}-\cite{6}-\cite{7}.

Such way of thinking doesn't take into account all the classical
symmetries enjoyed by these models, like the invariance under
diffeomorphisms, that the $BRS$ quantization breaks in an explicit
way. We have searched for another type ( non-standard ) of $BRS$
quantization which can take into account the classical
diffeomorphisms. We are going to show in this letter that it is
possible to quantize the classical model with a new type of gauge
fixing such that the quantum action is invariant under an extended
$BRS$ symmetry, obtained as a sum of the standard piece due to
gauge invariance and another one due to the diffeomorphisms.

\section{The classical $BF$ model}

The action of the $BF$ model in two dimensions is given by

\beq S_{inv} ( A, \phi ) = - \frac{1}{2} Tr \ \int \ d^2x \
\epsilon^{\mu\nu} F_{\mu\nu} \ \phi \label{21}\eeq

where $\epsilon^{\mu\nu}$ is the completely antisymmetric tensor,
$F_{\mu\nu} = F_{\mu\nu}^a T^a = \partial_\mu A_\nu -
\partial_\nu A_\mu - i [ A_\mu , A_\nu ]$ is the field strength
$ A_\mu (x) = A_\mu^a (x) T^a $ and $ \phi(x) = \phi^a(x) T^a $
are respectively the gauge field and the scalar field, both
belonging to the adjoint representation of the gauge group.

The action (\ref{21}) is classically invariant :

-i) under the gauge transformations, whose infinitesimal parameter
is $\Lambda (x) = \Lambda^a (x) T^a$

\begin{eqnarray}
\ & \ \delta_g A_\mu = D_\mu \Lambda = \partial_\mu \Lambda - i [
A_\mu , \Lambda ] \nonumber \\
\ & \ \delta_g \phi = - i [ \phi, \Lambda ] \label{22}
\end{eqnarray}

-ii) under the classical diffeomorphims with $\xi^\mu (x)$ as an
infinitesimal parameter

\begin{eqnarray}
\ & \ \delta_{\xi} A_\mu = \xi^\nu \partial_\nu A_\mu + (
\partial_\mu \xi^\nu ) A_\nu \nonumber \\
\ & \ \delta_g \phi = \xi^\nu \partial_\nu \phi \label{23}
\end{eqnarray}

Usually the quantization of the model is obtained with the
introduction of quantum fields ( $ c^a (x), \overline{c}^a (x),
b^a (x) $ ) playing the role respectively of ghost, antighost and
Lagrange multiplier.

In terms of those fields the gauge fixing reads in the Landau
gauge:

\beq S_{gh} = \int d^2 x \ \left( \ b^a \partial^\mu A_\mu^a -
\overline{c}^a \partial^\mu ( D_\mu c )^a \ \right) \label{24}
\eeq

The usual gauge transformations are no more a symmetry of the
gauge fixed quantum action $ S = S_{inv} ( A, \phi ) + S_{gf} ( A,
c, \overline{c}, b ) $, which is however invariant under the $BRS$
transformations:

\begin{eqnarray}
& \ & s_c A_\mu = D_\mu c \nonumber \\
& \ & s_c \phi = - i [ \phi, c] \nonumber \\
& \ & s_c c = i c^2 \nonumber \\
& \ & s_c \overline{c} = b \nonumber \\
& \ & s_c b = 0 \label{25}
\end{eqnarray}

The $BRS$ operator $s_c$ is characterized by two fundamental
properties:

- i) being nilpotent $s^2_c = 0$

- ii) being a symmetry of the quantum action

\beq s_c S = s_c ( S_{inv} + S_{gf} ) = 0 \label{26} \eeq

The last property (\ref{26}) is easily verified once that it is
noticed that the gauge fixing $S_{gf}$ is $s_c$-exact:

\beq S_{gf} ( b, c, \overline{c}, A ) = s_c \int d^2 x \
\overline{c}^a
\partial^\mu A_\mu^a \label{27} \eeq

The topological action $S_{inv}$ doesn't depend from the metric of
the space-time $g_{\mu\nu}$. In other words, only the gauge fixing
term of the action gives contribution to the energy-momentum
tensor, which is therefore an exact $BRS$ cocycle:

\beq T_{\mu\nu} = \frac{\partial S}{g^{\mu\nu}} = s_c \
\Lambda_{\mu\nu} \label{28} \eeq

The non trivial observation that both $T_{\mu\nu}$ and
$\Lambda_{\mu\nu}$ are conserved is the signal of an additional
symmetry of the action. The conservation law

\beq \partial^\nu \Lambda_{\mu\nu} = {\rm \ contact \ terms }
\label{29} \eeq

once integrated represents directly the Ward identity of the
vector supersymmetry:

\begin{eqnarray}
& \ & \delta_\mu A_\nu = 0 \nonumber \\
& \ & \delta_\mu \phi = \epsilon_{\mu\nu} \partial^{\nu} \overline{c} \nonumber \\
& \ & \delta_\mu c = - A_\mu \nonumber \\
& \ & \delta_\mu \overline{c} = 0 \nonumber \\
& \ & \delta_\mu b = \partial_\mu \overline{c} \label{210}
\end{eqnarray}

The existence of such linear vector symmetry is peculiar to the
topological field theories. It satisfies the following on-shell
algebra:

\begin{eqnarray}
& \ & s^2_c = 0 \nonumber \\
& \ & \{ \delta_\mu, \delta_\nu \} = 0 \nonumber \\
& \ & \{ \delta_\mu, s_c \} = \partial_\mu \label{211}
\end{eqnarray}

Our point of view is observing that the classical action
(\ref{21}) is invariant under infinitesimal diffeomorphisms and
that such invariance is ruined by the presence of the gauge
fixing.

We have searched to modify the gauge fixing in order to
incorporate such an invariance at a quantum level. The answer
turned out to be positive introducing an additional $BRS$ operator
for the diffeomorphisms \cite{8} :

\begin{eqnarray}
& \ & s_\xi \xi^\mu = \xi^\nu \partial_\nu \xi^\mu \nonumber \\
& \ & s_\xi \phi = \xi^\nu \partial_\nu \phi \nonumber \\
& \ & s_\xi A_\mu = \xi^\nu \partial_\nu A_\mu + ( \partial_\mu
\xi^\nu ) A_\nu \label{212}
\end{eqnarray}

with $\xi^\mu (x)$ a new field which we call ghost for the
diffeomorphisms, that shall be treated as an anticommuting
variable. Such $BRS$ operator $s_\xi$ enjoys the property of
nilpotency:

\beq s^2_\xi = 0 \ \ \ \ \leftrightarrow \ \ \ \ \ \ \{ \xi^\mu,
\xi^\nu \} = 0 \label{213} \eeq

It remains to redefine the quantum action

\beq \widetilde{S} = S_{inv} ( A, \phi ) + \widetilde{S}_{gf} ( A,
b, c, \overline{c}, \xi ) \label{214} \eeq

in order that it is totally $BRS$ invariant. In order to reach
such aim we define the new gauge fixing as :

\begin{eqnarray}
& \ & \widetilde{S}_{gf} ( A, b, c, \overline{c}, \xi ) = s \
\int d^2 x \ \overline{c}^a \partial^\mu A_\mu^a \nonumber \\
& \ & s = s_c + s_\xi \label{215}
\end{eqnarray}

where $ s = s_c + s_\xi $ is the total $BRS$ operator. The
modified action (\ref{215}) is still invariant under the vector
supersymmetry (\ref{210}).

 Having
defined the action of $s_\xi$ under the gauge ghost as:

\begin{eqnarray}
& \ & s_\xi \overline{c} = \xi^\mu \partial_\mu \overline{c}
\nonumber \\
& \ & s_\xi A_\mu = \xi^\nu \partial_\nu A_\mu + ( \partial_\mu
\xi^\nu ) A_\nu \label{216}
\end{eqnarray}

it is not difficult to isolate the contribution to the quantum
action due to the ghost of diffeomorphisms

\begin{eqnarray}
& \ & S_\xi = \int d^2 x \ \xi^\nu \ [ \ \partial_\nu \overline{c}
\ \partial^\mu A_\mu + \partial_\mu \overline{c} \ ( \partial_\mu
A_\nu - \partial_\nu A_\mu ) + \partial_\mu \partial_\mu
\overline{c} \ A_\nu ] = \nonumber \\
& \ & = \int d^2x \ \xi^\nu \ \delta_\nu \ L ( A, \phi, b, c,
\overline{c} ) \ \ \ \ \ \ \ \ S = \int d^2 x \ L \label{217}
\end{eqnarray}

We believe that it is not a coincidence that this contribution is
proportional to the variation of the Lagrangian under the vector
supersymmetry already met before ( eq. (\ref{210})), with the
ghost of diffeomorphisms playing the role of Lagrange multiplier.

\section{Chern-Simons action}

The Chern-Simons theory, once quantized in the usual manner in the
Landau gauge, leads to the following action \cite{9} :

\begin{eqnarray}
& \ & S = S_{inv} ( A ) + S_{gf} ( A, c, \overline{c}, b ) =
\nonumber \\
& \ & \int d^3 x \ Tr \ \left[ \ - \frac{1}{2}
\epsilon^{\mu\nu\rho} ( A_\mu \partial_\nu A_\rho + \frac{1}{3}
A_\mu [ A_\nu, A_\rho ]) + b \partial_\mu A^\mu - \overline{c}
\partial^\mu D_\mu c \ \right] \label{31}
\end{eqnarray}

and the associated $BRS$ transformations take the following form

\begin{eqnarray}
& \ & s^1_c A_\mu = D_\mu c \nonumber \\
& \ & s^1_c c = c^2 \nonumber \\
& \ & s^1_c \overline{c} = b \nonumber \\
& \ & s^1_c b = 0 \label{32}
\end{eqnarray}

The topological nature of this theory is ensured by an additional
vector supersymmetry

\begin{eqnarray}
& \ & \delta^1_\rho A_\mu = \epsilon_{\mu\rho\nu} \partial^\nu c \nonumber \\
& \ & \delta^1_\rho c = 0 \nonumber \\
& \ & \delta^1_\rho \overline{c} = A_\rho \nonumber \\
& \ & \delta^1_\rho b = D_\rho c \label{33}
\end{eqnarray}

It has been noticed in \cite{9} that this theory admits another
anti-$BRS$ symmetry, obtained by replacing the fields in this way

\beq A_\mu \rightarrow A_\mu, \ c \rightarrow \overline{c}, \
\overline{c} \rightarrow - c, \ b \rightarrow b - \{ c,
\overline{c} \} \label{34} \eeq

and this extends the $BRS-SUSY$ symmetry to an $N=2$ superalgebra.
By indicating the anti-$BRS$ and anti-$SUSY$ transformations with
$s^2_c$ and $\delta^2_\mu$, the on-shell algebra takes the form:

\beq \{ \delta^i_\mu, \delta^j_\nu \} = \epsilon^{ij}
\epsilon_{\mu\nu\tau} \partial^\tau \ \ \ \ \ \ \ \{ s^i_c,
\delta^j_\mu \} = \epsilon^{ij}
\partial_\mu \label{35} \eeq

Again our trick works by adding to the standard gauge-fixing of
the Landau gauge the contribution of the ghost of diffeomorphisms:

\begin{eqnarray}
& \ & \widetilde{S}_{gf} ( A, b, c, \overline{c}, \xi ) = s \int
d^3 x
\ \overline{c}^a \partial^\mu A_\mu^a \nonumber \\
& \ & s = s^1_c + s_\xi \label{36}
\end{eqnarray}

The part of the action containing the ghost of the diffeomorphisms
is again strictly related to the vector supersymmetry of type
$\delta^2$:

\beq S_\xi = \int d^3 x \ \xi^\mu \delta^2_\mu L( A, b, c,
\overline{c} ) \ \ \ \ \ \ S = \int d^3 x \ L \label{37} \eeq

\section{Conclusion}

We have seen in this letter that topological field theories in the
Landau gauge enjoy many interesting properties. We have found a
new choice of the gauge fixing such that the quantum model is
invariant under an extended $BRS$ symmetry including the vector
supersymmetry in a natural way. We hope that our suggestion may
help in elaborating further developments of the quantum properties
of these models.

\end{document}